\documentclass[showpacs,preprintnumbers,amsmath,amssymb,twocolumn,showkeys]{revtex4}
\usepackage{graphicx}   
\usepackage{dcolumn}    
\usepackage{bm}         
\usepackage{ifpdf}
\usepackage{graphicx,amssymb,lineno}
\usepackage{amssymb,lineno,amsfonts}
\usepackage{graphics}   
\usepackage{dcolumn}    
\usepackage{bm}         
\usepackage{bbm}
\usepackage{mathrsfs}
\usepackage{upgreek}
\usepackage{mathtools}

\begin{document}

\title{Dynamical Decoupling of Spin-Clusters using Solid State NMR}

\author{Abhishek Shukla and T. S. Mahesh}
\email{mahesh.ts@iiserpune.ac.in}
\affiliation{NMR Research Center,
Indian Institute of Science Education and Research (IISER), Pune 411008, India}

\date{\today}

\begin{abstract}
{
In this work we experimentally study the efficiency of various dynamical decoupling
sequences for suppressing decoherence of single as well as multiple
quantum coherences on large spin-clusters.  The system involves crystallites
of a powdered sample containing a large number of molecular protons interacting
via long-range dipole-dipole interaction. The multiple quantum coherences are prepared by
progressively creating correlations in the spin lattice
using standard pulse sequences implementing two-quantum average Hamiltonian.  
The spin system is then subjected to various dynamical decoupling sequences, followed by
conversion into observable single quantum coherence by using time-reversal sequence.
The experiments reveal superior performance of the recently introduced RUDD sequences
in suppressing the decoherence.
}      
\end{abstract}

\maketitle

\section{Introduction}
The study of dynamics and control of quantum many body systems has 
renewed interest in the field of quantum information.  While encoding
information onto a quantum channel can potentially speed up certain 
computations and allow secure data transmission, the practical realization
of these applications are hindered by the extreme sensitivity of the
quantum channel to environmental noises.
Systems based on nuclear spin-clusters is one among the various 
architectures being investigated to realize quantum channels.
Several experimental demonstrations of quantum information processing (QIP)
using solid-state nuclear magnetic resonance (SSNMR) have already been reported
\cite{ding,Mueller,laflammessnmr,moussa,Morton}.
By sophisticated control of spin-dynamics it is in principle possible to achieve 
a larger number of quantum bits (qubits) using SSNMR, because of the 
availability of large spin-clusters coupled
mutually through long-range dipole-dipole interactions.  
However in such a spin-cluster, fluctuating local fields at the site of each
spin induced by its environment leads to the decoherence of the encoded quantum 
information.

Due to the availability of large spin-clusters it is possible to prepare coherences
of large quantum numbers by a widening network of correlated spins evolving under
two-quantum average Hamiltonian \cite{pines83,pines85}.  These higher order coherences
are not directly observable as macroscopic magnetizations, but can be converted
into observable single quantum coherence (SQC) using a time-reversed two-quantum
average Hamiltonian.  This method, often known as a `spin-counting experiment'
has been used to study the evolution of coherences of large quantum numbers exceeding 
4000 \cite{suter2006,suter2007,suter2010a}.

Under the standard Zeeman Hamiltonian any spin-coherence is a non-equilibrium state 
and decays via various relaxation processes, ultimately leading to the equilibrium
longitudinal magnetization.  It has long been discovered that the decay process
of the spin coherence can be prolonged by applying a series of spin flips at regular
intervals of time.  This sequence known as `CPMG sequence', not only refocuses the
effect of spectrometer inhomogeneities, but also reverses the phase evolution of
the coherences under the random fluctuations, provided the spin flips are applied 
sufficiently frequent \cite{cp,mg}.  Such a dynamical method for the suppression 
of decoherence of a qubit due to its interaction with environment is often termed 
as `dynamical decoupling' (DD) \cite{lloyd}.
Recently Uhrig introduced a non-periodic spin-flip sequence which he proved
theoretically to provide optimal decoupling performance for dephasing spin-bath
interactions \cite{uhrig07}.  
CPMG and other similar periodic spin-flip sequences suppress spin-environment 
interaction to $n$th order using up to $O(2^n)$ pulses, while Uhrig dynamical 
decoupling (UDD) suppresses the same using only $n$ pulses.  
Filter function analysis of dynamical decoupling indicates that,
in a high frequency dominated bath with a sharp cutoff,
UDD works well provided the frequency of the spin-flips exceeds the 
cutoff frequency \cite{Biercuk,BiercukPRA,biercukjpb}.
On the other hand when the spectral density of the
bath has a soft cutoff (such as a broad Gaussian or Lorentzian),
the CPMG sequence was found to outperform the UDD
sequence \cite{duuhrig,dieteruhrig1,Cywinsky,Lange,Barthel,Ryan,sagi,Suter2011}. 
Later on, UDD has been generalized to suppress simultaneously both transverse
dephasing and longitudinal relaxation of a qubit \cite{lidar}.
Also, the original sequence for UDD is based the assumption of instantaneous 
spin-flips, which requires infinite bandwidth.  
More recently, Uhrig provided an improved sequence - `realistic UDD' (RUDD) for 
practical implementations with finite bandwidth \cite{rudd}.   

Most of the theory and experiments of DD sequences are for single spin
systems. Du et al have studied DD of electron spin coherence in solids 
\cite{duuhrig},
while Suter and co-workers have reported systematic experimental comparisons
of various DD schemes on an ensemble of single spins in SSNMR 
\cite{dieteruhrig1,Suter2011}.  

Later Agarwal has shown using theoretical and numerical calculations that even 
entangled states of two-qubit systems can be stored more efficiently
using UDD \cite{agarwal}.
 Similar theoretical studies were also carried out by Mukhtar et. al. \cite{Mukhtar}.
Experimentally, Wang et. al. have studied DD on electron-nuclear spin pairs 
in a solid-state system \cite{chinese2qudd}, and 
Soumya et. al. have studied the performance of UDD an a two-qubit liquid-state NMR system
\cite{soumyadd}.

In this article, we report the experimental study of performance of 
various DD schemes on an extended network of spin-1/2 nuclei forming a
large spin cluster.  This article is organized as
follows. Next section briefly describes the method of preparing
and detecting multiple quantum coherences (MQC) in SSNMR,  section III
summarizes the construction of various DD sequences, and the experimental
details are described in section IV. Finally we conclude in section V.

\section{Multiple Quantum SSNMR} 
\label{mqddsec}
The SSNMR Hamiltonian for a spin cluster with $M$ spin-1/2 nuclei is
\begin{equation}
{\cal H}_{int}= {\cal H}_Z+ {\cal H}_D,
\end{equation}
where the Zeeman and the secular part of dipolar interaction are
respectively,
\begin{eqnarray}
{\cal H}_Z  = \sum_{i=1}^{M} \omega_{i} I_z^i, \nonumber \\
{\cal H}_D = \sum_{i < j}D_{ij}\left[3I_z^i I_z^j-{\bf I}^i\cdot{\bf I}^j\right]
\end{eqnarray}
\cite{slichter}.
Here ${\bf I}^i$ and $I_z^i$ are spin angular momentum operator and its
z-component corresponding to the $i^{th}$ spin, and $w_i$ and $D_{ij}$ are
the chemical shift and the dipolar coupling constants.  
The equilibrium density matrix for the above Hamiltonian corresponds to
the longitudinal magnetization expressed as $\sum_i I_z^i$.
The density matrices for the longitudinal spin order can be expressed using
product of longitudinal spin operators, eg. $I_z^1 I_z^2 \cdots $.  
The coherences are described by the product of transverse 
(or of transverse and longitudinal) spin operators, eg. $I_z^1 I_x^2 I_x^3 \cdots $.
The transverse spin operators can also be expressed in terms of raising and lowering
operators:  $I_x = (I_+ + I_-)/2$ and $I_y = -i(I_+ - I_-)/2$.  The difference
between the total number of raising and lowering operators gives the quantum number
$n$ of a particular coherence.  For example, operators $I_+^j I_-^k$, $I_+^j$, and $I_+^j I_+^k$
describe zero, single, and two-quantum coherences respectively \cite{LevBook}.

\begin{figure}
\begin{center}
\includegraphics[width=8cm]{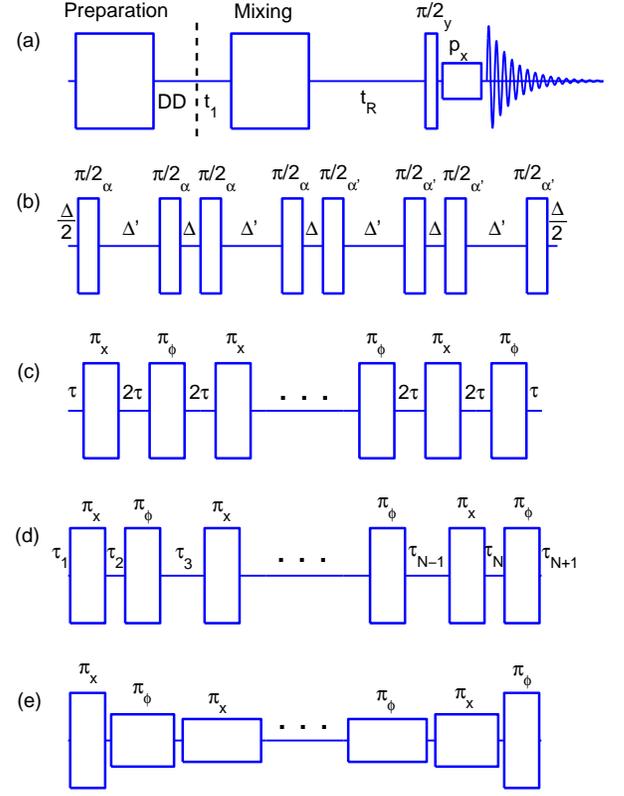}
\end{center}
\caption{\label{mqdd}
The experimental scheme (a) for studying the performance of DD on
large spin-clusters and the 8-pulse sequence (b) implementing ${\cal H}_1(\alpha)$.
In (b), $\Delta' = 2\Delta+\tau_{\pi/2}$, where $\tau_{\pi/2}$
is the duration of each $\pi/2$ pulse.  The DD schemes are described
in (c-e). The phase $\phi$ is set to $x$ for CPMG, UDD, and RUDD 
schemes, while it is alternated between $x$ and $-x$ for CPMGp, UDDp,
and RUDDp.}
\end{figure}

\begin{figure}
\begin{center}
\includegraphics[width=5cm]{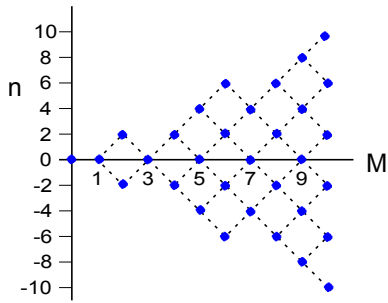}
\end{center}
\caption{\label{nVsM}
Possible quantum numbers ($n$) for $M$-spin cluster after excitation with 
several cycles of 8-pulse sequence (Fig. \ref{mqdd}b) implementing 
two-quantum average Hamiltonian $({\cal H}_m)$.
}
\end{figure}

The pulse sequence for preparing and detecting MQC
is shown in Fig. \ref{mqdd}(a-b).  The sequence in Fig. \ref{mqdd}a involves preparation
of MQC, application of DD schemes, free-evolution ($t_1$),
converting MQC into longitudinal spin order (mixing), 
destroying the residual coherences by transverse relaxation ($t_R$),  followed 
by detection after converting the longitudinal spin order into SQC.
The 8-pulse sequence in Fig. \ref{mqdd}b corresponds to the two-quantum average Hamiltonian
\begin{equation}
{\cal H}_1 = \frac{D_{ij}}{2} \left( I_+^i I_+^j + I_-^i I_-^j \right),
\end{equation}
for $\alpha=0$.
Preparation and mixing parts involve $m$-cycles of the 8-pulse sequence
${\cal H}_m(\alpha)$ and ${\cal H}_m(\pi/2)$ \cite{pines85}.
Under the preparation sequence, each uncorrelated spin gets correlated
with its neighboring spins, creating two-quantum coherence.  Each pair
of correlated spins evolve under their neighbors, and this growing
network of correlated spins in the form of higher order coherences
leads to large correlated clusters.
The possible quantum numbers and the corresponding cluster size increases
with the number of cycles.  Only even quantum coherences are prepared
as shown in Fig. \ref{nVsM}.  However, MQCs themselves have no detectable
macroscopic observables.  MQCs can be converted back to longitudinal spin
order by a time-reversal sequence ${\cal H}_m(\pi/2)$ (mixing), and subsequently
be converted into detectable SQCs by a $\pi/2_y$ pulse.
To separate the MQCs, the relative phase 
$\alpha$ between the preparation and mixing is incremented in proportion 
to the evolution time $t_1$ (Fig. \ref{mqdd}a).  Spurious transverse coherences are
suppressed by an extended delay $t_R$.  
A final purge pulse $p_x$ is used to keep only the 
$x$-component.

\section{DD schemes}
\subsection{CPMG and CPMGp}
\label{cpmg}
CPMG and CPMGp schemes involve periodic spin flips as shown in Figure \ref{mqdd}(c).
The phase $\phi$ is set to $x$ in CPMG, while it is alternated between $x$ and $-x$ in
CPMGp.  
CPMG and CPMGp have different performances depending on
the initial states \cite{dieteruhrig1,barrett08a,barrett08b}.  The total duration of the N-pulse 
CPMG is $T = N(2\tau+\tau_\pi)$, where $2\tau$ is the delay between the $\pi$ pulses and
$\tau_\pi$ is the duration of each $\pi$ pulse.
The same parameters $N$ and $T$ are used to compare
CPMG with the following schemes.

\subsection{UDD and UDDp}
The pulse distributions for UDD and UDDp schemes are shown in Fig. \ref{mqdd}(d).
Here the spin flips are symmetric but not periodic \cite{uhrig07}.  
The $j^{th}$ $\pi$ pulse is applied at the time instant
\begin{eqnarray}
t_j= T \sin^{2} \left[ \frac{\pi j}{2N+2} \right],
\label{uddtj}
\end{eqnarray}
where $T$ is the total duration of the sequence and $N$ is the total number of pulses.
For a finite bandwidth case, with a $\pi$ pulse
of duration $\tau_\pi$, the delays $\tau_j$ are given by
$\tau_1 = \tau_{N+1} = t_1-\tau_\pi/2$,
$\tau_j = t_{j+1}-t_j-\tau_\pi$, for $2 \le j \le N$.
Like in the previous scheme, UDD and UDDp are differed by the constant phase
and the phase alternation in $\phi$.

\subsection{RUDD}
\label{rudd}
In RUDD and RUDDp, in addition to the delays between the pulses, the pulse durations 
and amplitudes also vary, but the
overall sequence remains symmetric \cite{rudd}.  The pulse durations are given by
\begin{equation}
\tau_\pi^j = T \left[ \sin \left( \frac{\pi j}{N+1} \right) \sin \theta_p \right],
\end{equation}
where $T$ is total duration of the sequence and $N$ is the number of pulses.
Here $\theta_p$ is a constant and can be determined by the allowed bandwidth.
We choose $\tau_\pi^1 =\tau_\pi$, and calculated $\theta_p$ based on the minimum allowed pulse duration:
\begin{equation}
\mathcal
\sin\theta_p = \frac{\tau_\pi}{T \sin \left( \frac{\pi}{N+1} \right)}.
\end{equation}
The amplitude $a_j$ of $j^{th}$ pulse is calibrated such that $2\pi a_j\tau_\pi^j = \pi$.
Time instants of the center of each pulse is same as in equation (\ref{uddtj}).  Using these
time instants, the delays between the pulses can be calculated as
$\tau_{1} = \tau_{N+1} = t_1-\tau_\pi/2$, and
$\tau_{j} = t_j-t_{j-1}-\tau_\pi^j/2-\tau_\pi^{j-1}/2$ for $2 \le j \le N$.
Like in the previous schemes, RUDD and RUDDp are differed by the constant phase
and the phase alternation in $\phi$.

\section{Experiment}
The sample consists of crystallites of powdered hexamethylbenzene. At room temperature
the entire molecule undergoes six fold hopping about the $C_{6}$ axis of benzene ring \cite{pines85}.
Further the methyl group rapidly reorients about its $C_{3}$ axis.  Due to these motions,
the intramolecular dipolar interactions are averaged out.  Intermolecular dipolar 
coupling is retained and each molecule acts as a point dipole \cite{pines85}.  Under free precession (no DD),
this sample has a spin-spin relaxation constant of about 25 $\mu$s and a spin-lattice 
relaxation constant of 1.7s.  All the experiments are carried out on a Bruker 500 MHz
spectrometer at room temperature.

\begin{figure}
\begin{center}
\hspace{-1cm}
\includegraphics[width=7cm,angle=-90]{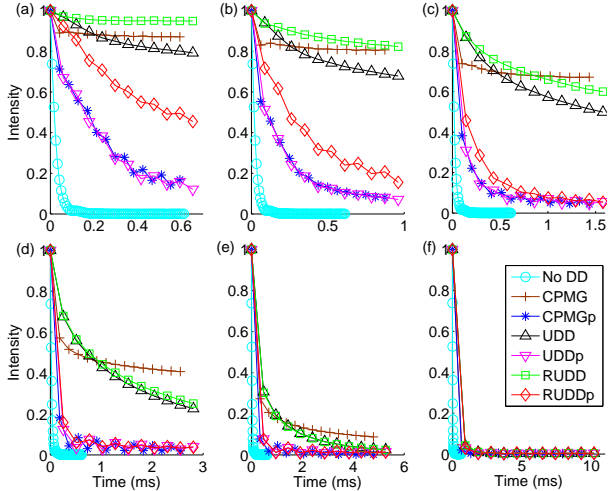}
\end{center}
\caption{\label{sqdddat} 
Performance of 7-pulse DD cycles on SQC.  The graphs correspond 
to $\tau=2\;\mu$s (a), $\tau=4\;\mu$s (b), $\tau=8\;\mu$s (c), $\tau=16\;\mu$s (d),
$\tau=32\;\mu$s (e), and $\tau=64\;\mu$s (f). 
} 
\end{figure}

\subsection{DD on SQC}
First we describe the performance of various DD schemes on SQC.
SQC was prepared by using an initial $(\pi/2)_y$ pulse
on equilibrium longitudinal magnetization.
As described in section \label{cpmg} and Fig. \ref{mqdd}c, CPMG sequences were 
constructed by periodic distribution of $\pi$ pulses in $\tau-\pi-\tau$ fashion.
The minimum $\tau$ in our experiments was set to 2 $\mu$s owing to the duty cycle
limit of the probe coil.  Also duration of $\pi$ pulse 
was found to be $\tau_\pi=4.3 \;\mu$s at maximum allowed amplitude.
Under these experimental conditions, the allowed values of $N$ for UDD and RUDD are 1 to 7.
For $N \ge 8$, one obtains negative delays between the pulses.  Therefore to study DD schemes for 
longer durations, we cycled these 7-pulse DD sequences.  
The 7-pulse CPMG has a total cycle time of $T(\tau) = 7(2\tau+\tau_\pi)$.  
The results of these experiments
are shown in Fig. \ref{sqdddat}.  
The graphs correspond 
to $\tau$ values ranging from $2\;\mu$s to $64\;\mu$s.  
The corresponding $T(\tau)$ values
are used to select the sampling points in no DD, as well as to construct other DD sequences. 
It is clear from these data that  RUDD displays superior
performance for shorter $\tau$ values, while CPMG shows better performance for longer $\tau$ values.
We can also notice from these plots that the performance of RUDDp is better than 
CPMGp and UDDp which have almost same performance.  UDD has better behavior than RUDDp, and 
for longer $\tau$ values UDD and RUDD have same behavior.
However, these performances in general may also be dependent on initial states \cite{dieteruhrig1}.
   
\subsection{DD on MQC}
\begin{figure}
\begin{center}
\hspace{-1.2cm}
\includegraphics[width=6.2cm,angle=-90]{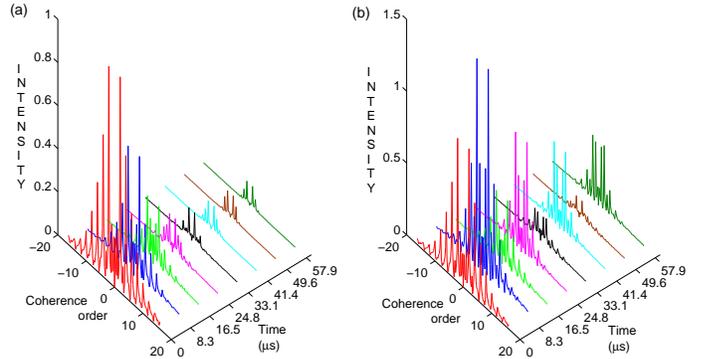}
\end{center}
\caption{\label{noddrudd} Multiple quantum spectra showing different coherence orders
detected after inserting different delays (a) and RUDD sequences of same durations (b).
}
\end{figure}
As described in section \ref{mqddsec} and Fig. \ref{mqdd}(a-b), the scheme for 
studying DD on MQC involves preparation of MQC, evolution of MQC, followed by 
storing MQC onto longitudinal spin-orders.
A delay $t_R = 5$ ms allows spurious coherences to decay.
The longitudinal spin order is then converted into SQC using a $(\pi/2)_y$
pulse, followed by a purge pulse $p_x$ of duration $50 \; \mu$s.  A 180 degree phase
alternation of the detection pulse, purge pulse, and the receiver is used to reduce
artifacts arising from receiver ringing \cite{pines85}.  For efficient generation of MQC
five cycles of 8-pulse sequence shown in Fig. \ref{mqdd}b was used in preparation and 
mixing periods, and the parameter $\Delta$ was optimized to $2 \;\mu$s.
In our experiments the coherences of successive quantum numbers are separated by 
$\Delta \omega = 2\pi \times 200$ kHz.
In order to separate a maximum of $n_\mathrm{max}$
coherences, the relative phase $\alpha$ between the preparation and mixing is incremented by
$\Delta \alpha = \pi/n_\mathrm{max}$, where
$n_\mathrm{max}$ was chosen to be 64.
The corresponding increment in the evolution period is
given by $\Delta t_1 = \Delta\alpha/\Delta\omega$.
The signal intensities of the spectrum corresponding to these increments after cosine transform
display strong peaks at even multiples of $\Delta \omega$.  
Mean value of the signal intensities is made to zero to suppress strong zero-quantum
peak.
Fig. \ref{noddrudd}a displays these even MQCs detected after inserting various delays,
and Fig. \ref{noddrudd}b displays those detected after applying RUDDp sequences of 
same durations. 
The first spectrum corresponding to no-delay is same in both cases, in which one
can easily observe MQCs of order up to 14.  Other spectra in (b)
were obtained by RUDDp sequences constructed with increasing number of pulses, i.e., 
$N={1,2,\cdots,7}$.
Under no DD (Fig. \ref{noddrudd}a), the intensities decay monotonically with delays,
while under RUDDp (Fig. \ref{noddrudd}b) the dependence of intensities is oscillatory w.r.t. 
$N$.  Similar behavior was earlier observed in a two-qubit liquid state
NMR system \cite{soumyadd}.  The spectra in (b) at odd $N$ clearly show better intensities 
compared to the corresponding spectra in (a). 
Comparisions of performance of different DD schemes for preserving MQCs of 
various orders are described in the following.

\begin{figure}
\label{mqdddat1}
\hspace{-1.1cm}
\includegraphics[width=6.4cm,angle=-90]{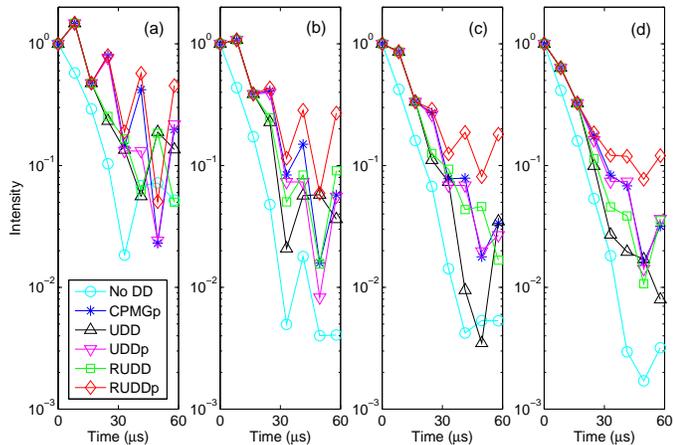}
\caption{\label{mqdddat1}Performance of various DD schemes in preserving MQCs of
order 2 (a), 4 (b), 6 (c), and 8 (d).  Each data set has 8 points, in which the
first point corresponds to no DD, and the rest correspond to different size of
the DD sequences with $N={1,2,\cdots,7}$.
}
\end{figure}
\begin{figure}
\hspace{-1.1cm}
\includegraphics[width=6.4cm,angle=-90]{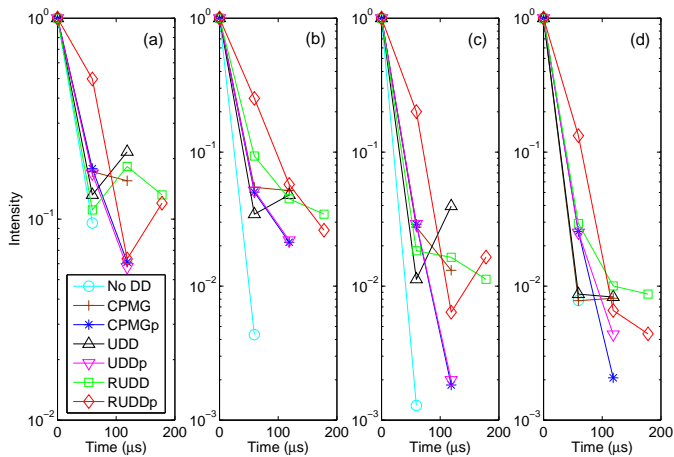}
\caption{\label{mqdddat2}
Performance of various DD schemes in preserving MQCs of
order 2 (a), 4 (b), 6 (c), and 8 (d).  The intensities were obtained
from spectra detected after applying up to a maximum of 3 cycles of 
7-pulse DD schemes. 
}
\end{figure}
The intensities of MQCs of even orders between 2 and 8 w.r.t. size of various DD 
schemes are plotted in Fig. \ref{mqdddat1}.  Although we can observe MQCs of
higher orders than 8 for short durations, for longer durations their intensity
measurements are not reliable due to noise. The first data point in each data set 
corresponds to no DD, and the rest correspond to different size of the DD sequence 
with $N={1,2,\cdots,7}$. As observed in Fig. \ref{noddrudd}b, we see the oscillatory
behavior of each MQC under various DD schemes.  But all the DD schemes display
an overall improvement w.r.t. no DD.  However it can be noticed that RUDDp has
significantly better performance than all other schemes, even for higher order 
coherences.  Surprisingly, unlike the single-quantum case, wherein RUDD displayed
the best performance, in multiple-quantum case RUDDp is the best scheme.

The intensities of MQCs of even orders between 2 and 8 for different
cycles of 7-pulse DD schemes are plotted in Fig. \ref{mqdddat2}. 
 The first data point in each case corresponds to
no DD.  The fast decay of magnetization under no DD allowed to detect
intensities corresponding to a duration of only one cycle, while for 
RUDD and RUDDp, intensities up to 3 cycles could be detected.
              
\section{Conclusions}
We studied the performance of various DD schemes on nuclear spins with
long-range interactions using a solid state NMR system.  First we applied
these DD schemes on a single quantum coherence.  The experiments were
carried out for various 7-pulse DD schemes and for different 
delays ($\tau$) between the $\pi$ pulses.  The results clearly show that all the DD schemes
are able to preserve the single quantum coherence for longer durations 
of time compared to no DD.  However, for small delays between the $\pi$
pulses, RUDD showed the best performance.  For longer delays between 
the $\pi$ pulses, CPMG was better.  Then we prepared MQCs of even orders 
using multiple cycles of the well known 8-pulse sequence implementing a 
two-quantum average Hamiltonian.  The MQCs so prepared could be
detected using standard spin-counting type experiments.  
We found the  Various DD schemes
were inserted after the preparation of MQCs.  We studied the
performance of these DD sequences with different sizes.  The intensity
behavior under all the DD sequences were oscillatory.  DD sequences
with odd number of $\pi$ pulses showed much better performance than
those with even number of $\pi$ pulses. RUDDp sequence showed
the best performance over all other sequences.                                                         
Since much of the theoretical and experimental work on dynamical decoupling
is limited to one or two qubits, we hope that the present work will shed
light on the behavior of dynamical decoupling in systems with a large number 
of correlated qubits.
It may also be interesting to study the behavior of odd quantum coherences
under DD.  It is possible to excite MQCs of all orders (both even and odd) 
simultaneously using pulse sequences implementing single quantum average
Hamiltonians \cite{pines87}. 
                          
\acknowledgments
Authors acknowledge useful discussions with 
Prof. Anil Kumar,
Prof. G. S. Agarwal,
Dr. Vikram Athalye,
Dr. Karthik Gopalakrishnan,
and S. S. Roy.
The use of 500 MHz NMR spectrometer at NMR Research Center, IISER-Pune
is acknowledged.


\references
\bibitem{ding}
S. Ding, C. A. McDowell, C. Ye, M. Zhan, X. Zhu, K. Gao, X. Sun, X. Mao, and M. Liu,
Eur. Phys. J. B
{\bf 24}, 23 (2001).

\bibitem{Mueller}
G. M. Leskowitz, N. Ghaderi, R. A. Olsen, and L. J. Mueller,
J. Chem. Phys. 
{\bf 119}, 1643 (2003).

\bibitem{laflammessnmr}
J. Baugh, O.Moussa, A. Ryan, R. Laflamme, C. Ramnathan, T. F. Havel, and D .G. Cory 
Phys. Rev. A
{\bf 104}, 22305 (2006).

\bibitem{moussa}
O. Moussa et al, C. Ryan, D. Cory, and R. Laflamme,
Phys. Rev. Lett.
{\bf 104}, 160501 (2010).

\bibitem{Morton}
S. Simmons, R. M. Brown, H. Riemann, N. V. Abrosimov, P. Becker, H. Pohl,
M. L. W. Thewalt, K. M. Itoh, and J. J. L. Morton,
Nature 
{\bf 470}, 69 (2011).

\bibitem{pines83}
Y. Yen and A. Pines,
J. Chem. Phys.
{\bf 78}, 3579 (1983).

\bibitem{pines85}
J. Baum, M. Munowitz, A. N. Garroway, and A. Pines,
J. Chem. Phys.
{\bf 83}, 2015 (1985).

\bibitem{suter2006}
H. Krojanski and D. Suter,
Phys. Rev. A
{\bf 74}, 062319 (2006).

\bibitem{suter2007}
M. Lovric, H. Krojanski, and D. Suter,
Phys. Rev. A
{\bf 75}, 042305 (2007).

\bibitem{suter2010a}
G. A. Alvarez and D. Suter,
Phys. Rev. Lett. 
{\bf 104}, 230403 (2010).

\bibitem{cp}
H. Y. Carr and E. M. Purcell,
Phys. Rev. 
{\bf 94}, 630 (1954).

\bibitem{mg}
S. Meiboom and D. Gill,   
Rev. Sci. Instr. 
{\bf 29}, 688 (1958)

\bibitem{lloyd}
L. Viola, E. Knill, and Seth Lloyd,
Phys. Rev. Lett.
{\bf 82}, 2417 (1999).

\bibitem{uhrig07}
G. S. Uhrig,
Phys. Rev. Lett.
{\bf 98}, 100504 (2007).

\bibitem{Biercuk}
M. J. Biercuk, H. Uys, A. P. VanDevender, N. Shiga,
W. M. Itano, and J. J. Bollinger, 
Nature 
{\bf 458}, 996 (2009).

\bibitem{BiercukPRA}
M. J. Biercuk, H. Uys, A. P. VanDevender, N. Shiga,
W. M. Itano, and J. J. Bollinger, 
Phys. Rev. A 
{\bf 79}, 062324 (2009).

\bibitem{biercukjpb}
M.J. Biercuk, A.C. Doherty, and H. Uys, 
arXiv:1012.4262 (2010). Accepted to Journal of Physics B.

\bibitem{duuhrig}
J. Du, X. Rong, N. Zhao, Y. Wang, J. Yang, and R. B. Liu,
Nature 
{\bf 461}, 1265 (2009).

\bibitem{dieteruhrig1}
G. A. Alvarez, A. Ajoy, X. Peng, and D. Suter,
Phys. Rev. A
{\bf 82}, 042306 (2010).

\bibitem{Cywinsky}
L. Cywinski, R. M. Lutchyn, C. P. Nave, and S. Das Sarma, 
Phys. Rev. B 
{\bf 77}, 174509 (2008).

\bibitem{Lange}
G. de Lange, Z. H. Wang, D. Riste, V. V. Dobrovitski,
and R. Hanson, 
Science 
{\bf 330}, 60 (2010).

\bibitem{Barthel}
C. Barthel, J. Medford, C. M. Marcus, M. P. Hanson,
and A. C. Gossard, 
arXiv:1007.4255 (2010).

\bibitem{Ryan}
C. A. Ryan, J. S. Hodges, and D. G. Cory, 
Phys. Rev. Lett. 
{\bf 105}, 200402 (2010).

\bibitem{sagi}
Yoav Sagi, Ido Almog, and Nir Davidson,
Phys. Rev. Lett. 
{\bf 105}, 053201 (2010).

\bibitem{Suter2011}
A. Ajoy, G. A. Alvarez, and D. Suter,
Phys. Rev. A 
{\bf 83}, 032303 (2011).

\bibitem{lidar}
J. R. West, B. H. Fong, and D. A. Lidar,
Phys. Rev. Lett.
{\bf 104}, 130501 (2010).

\bibitem{rudd}
S. Pasini, P. Karbach, and G. S. Uhrig,
arxiv: 1009.2638v2, (2010).

\bibitem{agarwal}
G. S. Agarwal,
Phys. Scr. 
{\bf 82}, (2010) 038103.

\bibitem{Mukhtar}
M. Mukhtar, W. T. Soh, T. B. Saw, and J. Gong,
Phys. Rev. A 
{\bf 82}, 052338 (2010).

\bibitem{chinese2qudd}
Y. Wang, X. Rong, P. Feng, W. Xu, B. Chong, Ji-Hu Su, J. Gong, and J. Du,
Phys. Rev. Lett.
{\bf 106}, 040501 (2011).

\bibitem{soumyadd}
S. S. Roy and T. S. Mahesh,
Phys. Rev. A 
{\bf 82}, 052302 (2010).

\bibitem{slichter}
C. P. Slichter,
{\it Principles of Magnetic Resonance},
Third Edition,
Springer (1996).

\bibitem{LevBook}
M. H. Levitt, 
{\it Spin Dynamics},
J. Wiley and Sons Ltd., 
Chichester
(2002).

\bibitem{barrett08a}
Y. Dong, R. G. Ramos, D. Li, and S. E. Barrett,
Phys. Rev. Lett.
{\bf 100}, 247601 (2008).

\bibitem{barrett08b}
D. Li, Y. Dong, R. G. Ramos, J. D. Murray, K. MacLean, A. E. Dementyev, and S. E. Barrett,
Phys. Rev. B
{\bf 77}, 214306 (2008).

\bibitem{pines87}
D. Suter, S. B. Liu, J. Baum, and A. Pines,
Chem. Phys.
{\bf 114}, 103 (1987).
\end{document}